\documentclass[prl,twocolumn,showpacs,amsmath,amssymb,superscriptaddress]{revtex4-2}
\usepackage{bbm}
\usepackage{mathrsfs}
\usepackage{epsfig}
\usepackage{graphicx}
\usepackage{amsfonts}
\usepackage[figuresright]{rotating}
\usepackage{amssymb}
\usepackage{amsmath}
\usepackage{dcolumn}
\usepackage{bm}
\usepackage{braket}
\usepackage{comment}
\usepackage{mathtools}
\usepackage{xcolor}
\usepackage{multirow}
\usepackage{setspace}

\usepackage[colorlinks,linkcolor=blue,anchorcolor=blue,citecolor=blue,urlcolor=blue]{hyperref}
\usepackage{color}

\usepackage[normalem]{ulem}

\begin{document}
\title{Spontaneous magnetization in time-reversal symmetry-breaking unitary superconductors}

\author{Lun-Hui Hu}
\email{lvh5389@psu.edu}
\affiliation{Department of Physics, Pennsylvania State University, University Park, State College, PA 16802, USA}

\author{Xuepeng Wang}
\email{xpwang@smail.nju.edu.cn}
\affiliation{National Laboratory of Solid-State Microstructures, School of Physics, Nanjing University,	Nanjing 210093, China}

\author{T.\ Shang}
\email{tshang@phy.ecnu.edu.cn}
\affiliation{Key Laboratory of Polar Materials and Devices (MOE), School of Physics and Electronic Science, East China Normal University, Shanghai 200241, China}

\begin{abstract}
	We report the study of spontaneous magnetization (i.e., spin-polarization) for time-reversal symmetry (TRS)-breaking superconductors with unitary pairing potentials, in the absence of external magnetic fields or Zeeman fields.
	Spin-singlet ($\Delta_s$) and spin-triplet ($\Delta_t$) pairings can coexist in superconductors whose crystal structure lacks inversion symmetry. 
	The TRS can be spontaneously broken once a relative phase of $\pm\pi/2$ is developed, forming a TRS-breaking unitary pairing state ($\Delta_s\pm i\Delta_t$).
	We demonstrate that such unitary pairing could give rise to spontaneous spin-polarization with the help of spin-orbit coupling.
	Our result provides an alternative explanation to the TRS breaking, beyond the current understanding of such phenomena in the noncentrosymmetric superconductors.
	The experimental results of Zr$_3$Ir and CaPtAs are also discussed in the view of our theory.
\end{abstract}
\maketitle

\textit{Introduction--.}
In condensed matter physics, superconductivity and magnetism are generally antagonistic to each other \cite{sigrist1991,sigrist2009,smidman2017} and the interplay between them brings us intriguing phenomena.
One of them is the Fulde-Ferrell-Larkin–Ovchinnikov (FFLO) state \cite{fulde1964,larkin1965} in which superconducting Cooper pairs carry a finite momentum induced by an external Zeeman field.
Recent theoretical efforts have been made to realize chiral Majorana modes in the topological FF phase \cite{qu2013,zhang2013} and the Majorana mode chain in the topological LO phase \cite{hu2019a}.
Another fascinating phenomenon is the spontaneous magnetization or spin-polarization (SP) in a time-reversal symmetry (TRS)-breaking superconductor (SC) \cite{timm2015,yang2017,robins2018}, which continuously promotes extensive experimental and theoretical research \cite{wysokinski2019,ghosh2020a}. 
The TRS-breaking candidate SCs include Sr$_2$RuO$_4$ \cite{luke1998,xia2006,maeno2011}, Re$T$ ($T$ = transition metal) \cite{singh2014,pang2018,Shang2018a,Shang2018b,Shang2020ReMo,Shang2021}, UPt$_3$ \cite{luke1993,sauls1994,schemm2014}, UGe$_2$ \cite{saxena_nature_2000,huxley_prb_2001}, URhGe \cite{aoki_nature_2001},  UCoGe \cite{Huy2007,aoki_jpsj_2019},
PrOs$_4$Sb$_{12}$ \cite{aoki2003}, URu$_2$Si$_2$ \cite{mackenzie2003,schemm2015}, SrPtAs \cite{biswas2013}, Ru$_7$B$_3$ \cite{parzyk2014,cameron2019}, LaNiC$_2$ \cite{hillier2009}, LaNiGa$_2$ \cite{hillier2012,weng2016}, Bi/Ni bilayers \cite{gong2017}, CaPtAs \cite{Shang2020a}, Zr$_3$Ir \cite{Shang2020b}, and others summarized in a recent paper \cite{ghosh2020a}.
More recently, iron-based SCs also exhibit TRS-breaking signatures \cite{grinenko2020,zaki2019}.
These exciting experimental discoveries arouse considerable attentions, and a great deal of theoretical progress has recently been made on TRS-breaking SCs with mixed pairing states \cite{lee2009,wu2010,stanev2010,khodas2012,garaud2014,maiti2015,lin2016,wang2017,zhang2018,kang2018,wang2020}, non-unitary pairing states \cite{jorge2010,kallin2016,kozii2016,yuan2017,tkachov2017,mizushima2018,csire2018,brydon2019,hu2019b,lado2019,ghosh2020,yuan2020}, and Bogliubov Fermi surface \cite{agterberg2017,brydon2018,lapp2019,menke2019}.

There are mainly two direct ways to probe spontaneously TRS-breaking pairing states, including the zero-field muon-spin relaxation
($\mu$SR) \cite{schenck1985,lee1999,yaouanc2011}, and the polar Kerr effect (PKE) \cite{spielman1990,kapitulnik2009}.
Firstly, the $\mu$SR is especially very sensitive to a small change of internal fields (with a resolution down to 10 $\mu$T)~\cite{Amato1997}. 
The enhancement of the zero-field muon-spin depolarization rate in the superconducting state provides direct evidence for TRS breaking pairing states.
Besides, the PKE measures the optical phase difference between two opposite circular-polarized lights reflected on a sample
surface, thus it gives information about the TRS of a system.
A finite PKE unambiguously points to TRS-breaking states.
Theoretically, non-unitary spin-triplet pairing potentials, such as the $A_1$ phase in He$^3$ superfluid characterized by spin-triplet $\vec{d}_s(\mathbf{k})=k_z(1,-i,0)$~\cite{leggett_rmp_1975}, could spontaneously induce SP in a homogeneous SC and thus naturally explaining experimental observations by the $\mu$SR and the PKE \cite{sigrist1991,sigrist2009}.
However, one may still wonder if there is any mechanism other than the non-unitary spin-triplet pairing states to induce the SP in the TRS-breaking superconductors. 

In this work, we report the discovery of SP in TRS-breaking unitary SCs.
The spin-singlet pairing ($\Delta_s$) coexists with the spin-triplet pairing $(\Delta_t)$ in both noncentrosymmetric SCs and superconducting thin films. 
Once a relative phase of $\pm\pi/2$ is developed ($\Delta_s\pm i\Delta_t$), TRS is spontaneously broken.
By combing the symmetry analysis and the Ginzburg-Landau (GL) theory, we find that the interplay between spin-orbit coupling (SOC) and the $\Delta_s \pm i\Delta_t$ unitary pairing potential could give rise to SP in a homogeneous SC. 
The direction of the induced SP is perpendicular to both the SOC $\vec{g}$-vector and the spin-triplet $\vec{d}$-vector, even though both $\vec{g}$ and $\vec{d}$ are real vectors.
Our result provides an alternative explanation for the TRS breaking phenomenon. 
The potential applications of our theory to recently discovered noncentrosymmetric SCs (e.g. Zr$_3$Ir and CaPtAs) are also discussed.

\textit{TRS-breaking unitary pairings--.} 
In the absence of external magnetic fields or Zeeman fields, the non-vanishing magnetism or SP in the superconducting states generally causes the spontaneous breaking of TRS.
The SP can be generated by non-unitary pairing states with a complex spin-triplet $\vec{d}$-vector, whose direction is parallel to $\vec{d}\times \vec{d}^\ast$.
Alternatively, in this work, we explore the spontaneous SP induced by TRS-breaking unitary pairing states in SCs whose crystal structure lacks inversion symmetry.
For this propose, we start with a single-band model Hamiltonian \cite{frigeri2004},
\begin{align}\label{eq-ham0}
\mathcal{H}_0 = \xi_{k}\sigma_0 + \alpha \vec{g}\cdot\vec{\sigma}, 
\end{align}
where $\xi_k= k^2/2m-\mu$ is the electron band energy measured from the Fermi energy $\mu$, $\vec{\sigma}$ denotes the Pauli matrices of electron spin, and $\alpha$ is the strength of SOC.
The inversion symmetry is broken due to $\vec{g}(\mathbf{k}) = -\vec{g}(-\mathbf{k})$.
In this work, we mainly focus on the Rashba-type SOC which is given by $\vec{g}=(-k_y,k_x)$, allowed by the $C_{4v}$ point group.
There are two Fermi surfaces with opposite chirality, and the two bands are $\epsilon_{k,\pm}=\xi_k \pm \alpha \vert \vec{g}\vert$, and the Fermi momentum is $k_F=\sqrt{2m\mu}$ for $\alpha\to0$.

\begin{figure}[t]
	\centering
	\includegraphics[width=0.9\linewidth]{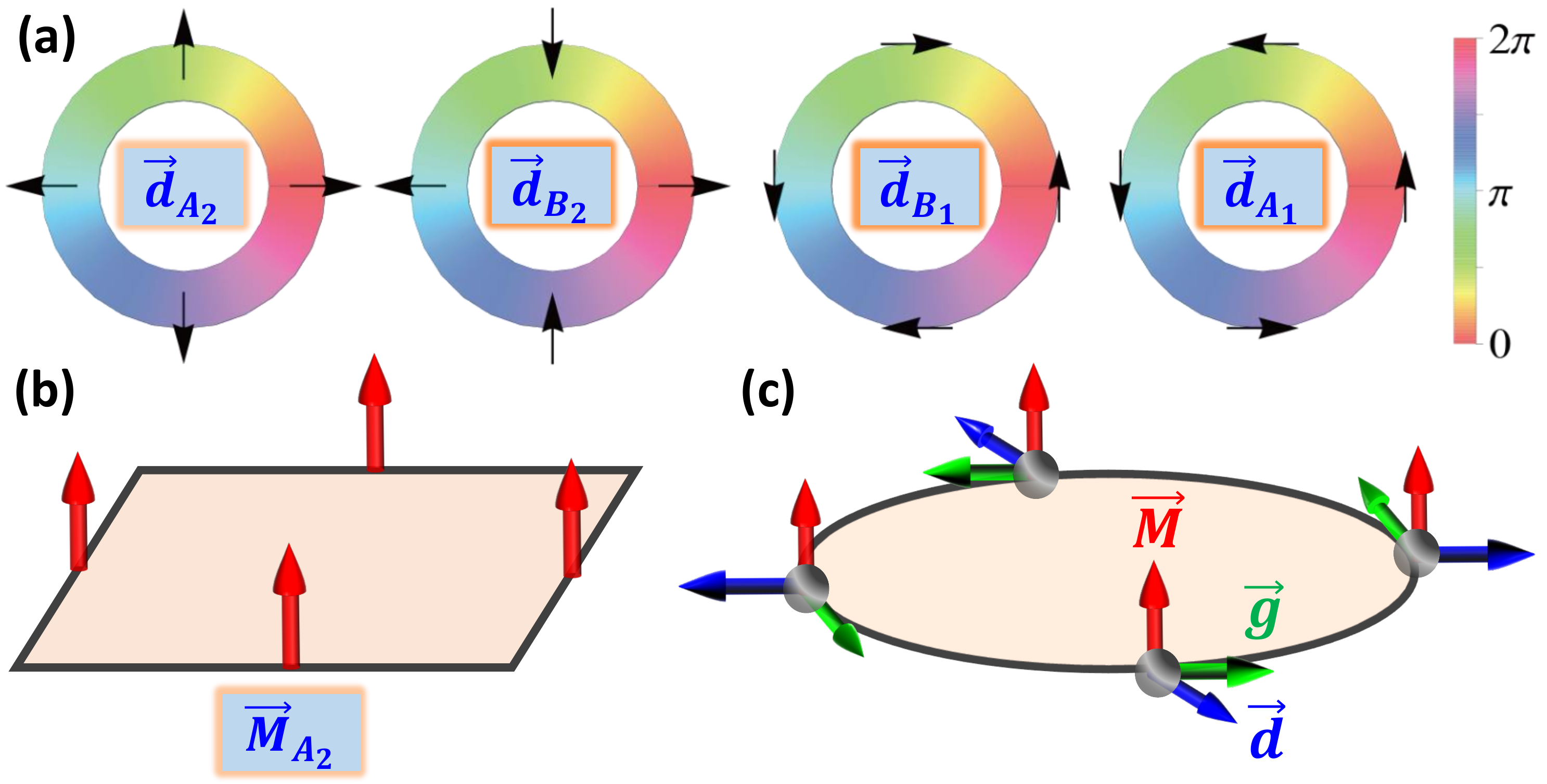}
	\caption{\label{fig1} Illustrations of the p-wave pairings and the boundary SP.
		(a) The four p-wave pairing symmetries with in-plane $\vec{d}$ vectors, labeled by the black arrows.
		(b) The real space viewpoint of the boundary SP of a two-dimensional rectangular sample, marked by the red arrows.
		(c) The momentum space viewpoint of the $\vec{M}_{A_{2}}$-type SP that is parallel with $\vec{g}\times\vec{d}$, because of the product rule of the point group: $A_2 = A_2\times A_1 (B_1\times B_2)$.
	}	
\end{figure}

Then we consider the superconducting pairing Hamiltonian with attractive interactions, 
\begin{align}
\mathcal{H}_{int} = \sum_{\mathbf{k},\mathbf{k}'}\sum_{s_1,s_2} V_{\mathbf{k},\mathbf{k}'} c_{\mathbf{k},s_1}^\dagger c_{-\mathbf{k},s_2}^\dagger c_{-\mathbf{k}',s_2} c_{\mathbf{k}',s_1},
\end{align}
here $s_{1,2}$ are spin indexes. 
Applying the mean-field decompositions, we define the gap functions as $\Delta_{s_1,s_2}(\mathbf{k}) = \sum_{\mathbf{k}'} V_{\mathbf{k},\mathbf{k}'} \left\langle c_{-\mathbf{k}',s_2} c_{\mathbf{k}',s_1} \right\rangle$.
Here $\langle\cdots\rangle$ represents averaging over the thermal equilibrium states.
After ignoring fluctuations, the mean-field pairing Hamiltonian becomes,
\begin{align}\label{eq-ham-delta}
\mathcal{H}_{\Delta} = \sum_{\mathbf{k}}\sum_{s_1,s_2} \Delta_{s_1,s_2}(\mathbf{k}) c_{\mathbf{k},s_1}^\dagger c_{-\mathbf{k},s_2}^\dagger + \text{h.c.}. 
\end{align}
Due to the breaking of inversion symmetry, the even-parity pairing coexists with the odd-parity pairing \cite{frigeri2004}. The pairing potential is generally given by $\hat{\Delta}(\mathbf{k})=( \Delta_s\psi(\mathbf{k}) + \Delta_t \vec{d}(\mathbf{k})\cdot\vec{\sigma} ) i\sigma_y$, with $\psi(\mathbf{k})=\psi(-\mathbf{k})$ and the real spin-triplet $\vec{d}$-vector $\vec{d}(\mathbf{k}) = -\vec{d}(-\mathbf{k})$ required by the Fermi statistic. 
And the pairing strengths $\Delta_{s(t)}=\vert \Delta_{s(t)}\vert\exp(i\theta_{s(t)})$ are generally complex for the spin-singlet (triplet) pairing states.
To break TRS $\mathcal{T}=i\sigma_y \mathcal{K}$  ($\mathcal{K}$ is a  complex conjugate), the relative phase $\theta_{ts}=\theta_t-\theta_s$ should be nonzero. 
Typically, $\theta_{ts}$ is pinned to $\pm\pi/2$, leading to the TRS breaking unitary pairing $\Delta_s\pm i\Delta_t$, 
because of $\hat{\Delta}(\mathbf{k})^\dagger \hat{\Delta}(\mathbf{k}) = \vert\Delta_s\psi(\mathbf{k})\vert^2 + \vert\Delta_t\vert^2\vec{d}(\mathbf{k})\cdot\vec{d}(\mathbf{k})$.
Here, the angular form factors for spin-singlet pairing could be either s-wave ($\psi_s(\mathbf{k})\sim \{1,\cos k_x +\cos k_y,\cos k_x\cos k_y\}$) or d-wave ($\psi_d(\mathbf{k})\sim \cos k_x - \cos k_y$).
As for p-wave pairing symmetries, we focus on in-plane $\vec{d}$ vectors for simplicity, and there are  four possibilities,
\begin{align}\label{eq-dvector}
\begin{split}
\vec{d}_{A_{1}}(\mathbf{k}) = (-\sin k_y,\sin k_x), \vec{d}_{A_{2}}(\mathbf{k}) = (\sin k_x,\sin k_y), \\
\vec{d}_{B_{1}}(\mathbf{k}) = (\sin k_y,\sin k_x), \vec{d}_{B_{2}}(\mathbf{k}) = (\sin k_x,-\sin k_y). 
\end{split}
\end{align}
They belong to different 1D irreducible representations of the $C_{4v}$ point group, shown in Fig.~\ref{fig1}(a). 
Note that all the spin-triplet $\vec{d}$-vectors in Eq.~\eqref{eq-dvector} have odd parity, and are real so that $\vert\vec{d}^\ast\times \vec{d} \vert^2$ vanishes.  
A relatively small SOC $\alpha k_F$ in the unit of $k_BT_c$ is assumed so that the spin-triplet $\vec{d}$-vectors that are not parallel with the SOC $\vec{g}$-vector could still be stabilized \cite{tanaka2007,ramires2018}.

\textit{Spontaneous spin-polarization--.}
Next, we use the GL theory to address the SP induced by the $\Delta_s\pm i \Delta_t$ unitary pairing potential in the absence of external magnetic fields or Zeeman fields.
The symmetry allowed free energy of a homogeneous SC reads,
\begin{align}\label{eq-GL-free-energy}
\mathcal{F} = F_2 + F_4 + \gamma_1 \left(\Delta_s^\ast \Delta_t\right)^2 + (\vec{\gamma_2}\cdot\vec{M}) \Delta_s^\ast\Delta_t  + \text{c.c.},
\end{align} 
where $F_2 = \alpha_s(T) \vert \Delta_s\vert ^2 +  \alpha_t(T) \vert \Delta_t\vert ^2 + \alpha_M\vert\vec{M}\vert^2$ and $F_4 = \beta_s \vert \Delta_s\vert ^4 +  \beta_t \vert \Delta_t\vert^4 + \beta_{st}\vert \Delta_s\vert^2\vert \Delta_t\vert^2$.
Here $\gamma_1\neq0$ indicates the pairing breaks TRS since the relative phase difference between singlet and triplet pairings is developed as $\pm \pi/2$ \cite{wang2017}. 
The bilinear coupling term $\Delta_s^\ast\Delta_t$ pines the phase difference to an arbitrary non-zero value \cite{hinojosa2014} in the low temperature. 
Hereafter, we consider both $\Delta_s$ and $\Delta_t$ belong to different representations of the lattice symmetry group thus the bilinear coupling is forbidden. 
Here $\alpha_M>0$ indicates the lacking of intrinsic ferromagnetic ordering, and the $\gamma_2$-terms couple the SP $\vec{M}$ with the TRS-breaking unitary pairing $\Delta_s\pm i \Delta_t$, which satisfy both the global $U(1)$ symmetry ($\theta_{s,t}\to \theta_{s,t} +\delta\theta$) and TRS. 
Once $\vec{\gamma}_2\neq 0$, $\vec{M}$ is spontaneously induced by unitary $\Delta_s\pm i\Delta_t$. 
By minimizing $\mathcal{F}$, we find 
\begin{align}
\vec{M}=-\tfrac{1}{\alpha_M}\text{Im}[\vec{\gamma}_2 \Delta_s^\ast\Delta_t],
\end{align} 
which will not alter the relative phase difference $\theta_{ts}$ between $\Delta_s$ and $\Delta_t$ (see Sec.~A in the Supplementary Materials). 
It indicates that the unitary pairing states $\Delta_s\pm i\Delta_t$ coexist with the induced SP $\vec{M}$.
The bulk magnetization vanishes in a purely clean system due to two reasons: the translational symmetry and the Meissner effect \cite{shopova2005,chirolli2017}. 
Nevertheless, the boundary SP (see Fig.~\ref{fig1}(b)) persists as well as the SP around impurities in the bulk, both of which can be detected by the $\mu$SR and the PKE.

Below we first use symmetry analysis to classify the $\vec{\gamma}_2$-vector.
The free energy in Eq.~\eqref{eq-GL-free-energy} preserves all the crystalline symmetries, which 
imposes strict constraints on the direction of the SP $\vec{M}$. 
We take the $C_{4v}$ point group as an example to identify the ferromagnetic-type SP induced by the TRS-breaking unitary pairing potential.
A full classification for the $\vec{\gamma}_2$-terms by different symmetry groups is left for future work. 
The $C_{4v}$ point group is generated by three independent symmetry operators ($\sigma_v$, $\sigma_d$ and $R_{4z}$):
$\sigma_v$ are the vertical reflection planes along $x$ and $y$; $\sigma_d$ are the diagonal reflection planes along the $x\pm y$ lines; $R_{4z}$ is the four-fold rotation along $z$ axis. 
To break both $\sigma_d$ and $\sigma_v$ simultaneously, only the z-component of $\vec{\gamma}_2$ is nonzero, namely, $\vec{\gamma}_2=(0,0,\gamma_{2}^z)$, illustrated in Fig.~\ref{fig1}(b). 
It belongs to $A_2$ representation of $C_{4v}$, shown in Table~(S1) in the Supplementary Materials.
According to the product rule of the point group, we conclude that the SP could be induced by the interplay of the $s$-wave ($d$-wave) singlet pairing and the $\vec{d}_{A_{2}}(\mathbf{k})$ [$\vec{d}_{B_{2}}(\mathbf{k})$]-triplet pairing, 
\begin{align}\label{eq-sip-and-dip}
\begin{split}
\hat{\Delta}_{s+ip} &= \left(\Delta_s \Psi_s(\mathbf{k}) + i\Delta_t \vec{d}_{A_{2}}(\mathbf{k})\cdot\vec{\sigma} \right)i\sigma_y,\\
\hat{\Delta}_{d+ip} &= \left(\Delta_s \Psi_d(\mathbf{k}) + i\Delta_t \vec{d}_{B_{2}}(\mathbf{k})\cdot\vec{\sigma} \right)i\sigma_y.
\end{split}
\end{align}
Both the $s+ip$- and $d+ip$-pairing states are fully gapped in 2D SCs, while gap nodes can exist in 3D SCs.

Next, we investigate the important role of Rashba SOC for the establishment of the $\vec{\gamma}_2$-term in Eq.~\eqref{eq-GL-free-energy}.
In this work, we study a single-band Hamiltonian with SOC in Eq.~\eqref{eq-ham0}, and the results could also be generalized to multi-band systems.
The coupling coefficients of $\vec{\gamma}_2$ are calculated for the Hamiltonian $\mathcal{H}_0+\mathcal{H}_\Delta$,
\begin{align}
\gamma_2^{i} = \frac{1}{\beta} \sum_{\mathbf{k},\omega_n} \text{Tr}\left\lbrack G_h (\vec{d}\cdot\vec{\sigma}) G_e \sigma_i G_e \sigma_0 \right\rbrack,
\end{align}
with $i=\{x,y,z\}$ and $\beta=1/(k_BT)$ the inverse of temperature.
And the Matsubara Green’s function is $G_e(\mathbf{k},i\omega_n) = [i\omega_n - \mathcal{H}_0(\mathbf{k})]^{-1}$ with $\omega_n=(2n+1)\pi/\beta$ and $G_h(\mathbf{k},i\omega_n)=G_e^\ast(-\mathbf{k},-i\omega_n)$.
We find that the direction of $\vec{M}$ is perpendicular to both $\vec{d}$ vector and the SOC $\vec{g}$ vector, namely,
\begin{align}
\vec{\gamma}_2 \propto i\sum_{\mathbf{k}}\left\langle \left(\vec{g}(\mathbf{k}) \times \vec{d}(\mathbf{k}) \right) \cdot \psi_s(\mathbf{k}) \right\rangle_{FS},
\end{align}
where $\langle \cdots \rangle_{FS}$ denotes the average over the entire Fermi surfaces.
Here we take the $\Delta_s+i\Delta_t$ with $s$-wave pairing and $\vec{d}_{A_{2}}(\mathbf{k})$ as an example [see Fig.~\ref{fig1}(c)], where 
$\vec{g}\times\vec{d}_{A_{2}} \approx \vec{e}_z (\sin^2 k_x+\sin^2 k_y)$.  
Moreover, we find the nonzero z-component of $\vec{\gamma}_2$ to the leading order of $\alpha k_F/k_BT_c$ as,
\begin{align}\label{eq-gamma2z-approx}
\gamma_2^z = \frac{7\zeta(3)}{8\pi^3}\frac{\alpha k_F}{k_BT_c},
\end{align}
where $\zeta(z)$ is the Riemann zeta function.
The crucial role of SOC to the SP is manifest in Eq.~\eqref{eq-gamma2z-approx}, representing the key result of this work. 
Only when $\alpha\neq0$, we can have $\gamma_2^z\neq0$ as well as $\vec{M}_z\neq0$. 
Therefore, SOC is indispensable to induce SP by a TRS-breaking unitary pairing $\Delta_s+i\Delta_t$.
Moreover, the sign of the SP is determined by sign$(M_z)$ = sign$(\alpha\Delta_t\Delta_s)$.

\begin{figure}[!htbp]
	\centering
	\includegraphics[width=0.75\linewidth]{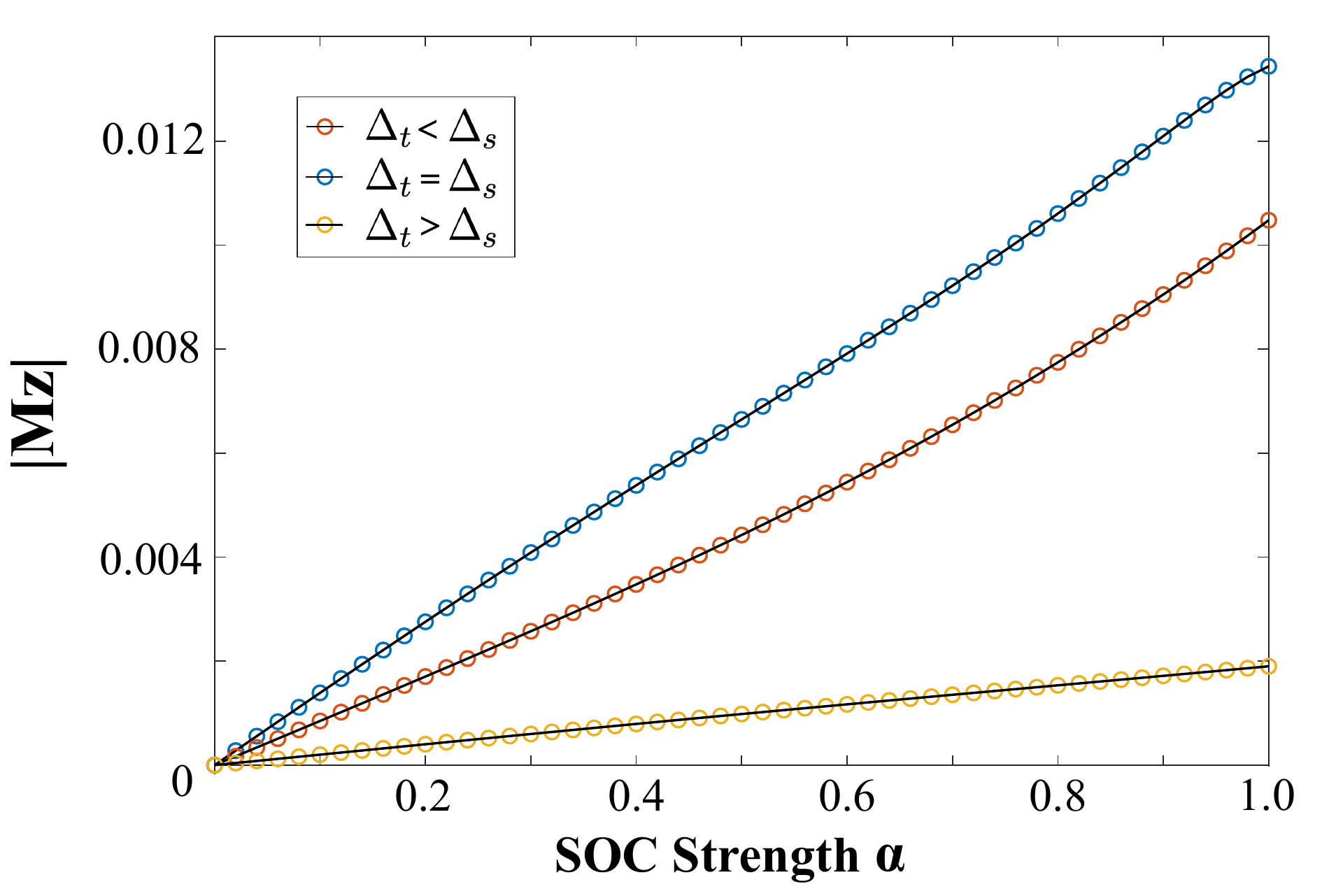}
	\caption{\label{fig2} The spontaneous SP induced by the $d+ip$ unitary pairing potential given by Eq.~\eqref{eq-sip-and-dip}. The averaged SP on the boundary is calculated as a function of SOC strength for three cases, $\Delta_t<\Delta_s$ (red circle), $\Delta_t=\Delta_s$ (blue circle) and $\Delta_t>\Delta_s$ (orange circle). 
		Parameters: $m_0=1$, $\mu=1$, and $(\Delta_t,\Delta_s)=\{(1,2),(1,1),(2,1)\}$ are used for the three cases.
	}	
\end{figure}

To be more explicitly, we perform a numerical calculation for the averaged SP based on the solution of the Bogoliubov–de Gennes (BdG) Hamiltonian,
\begin{align}
\mathcal{H}_{BdG}=\left(\begin{array}{cc}
\mathcal{H}_0(\mathbf{k})           & \mathcal{H}_\Delta \\ 
\mathcal{H}_\Delta^\dagger & -\mathcal{H}_0^\ast(-\mathbf{k})
\end{array} \right),
\end{align}
where the Nambu basis $(c_{\mathbf{k},\uparrow}^\dagger, c_{\mathbf{k},\downarrow}^\dagger, c_{-\mathbf{k},\uparrow},c_{-\mathbf{k},\downarrow})^T$ is used.
Here $\mathcal{H}_0(\mathbf{k})$ is given by Eq.~\eqref{eq-ham0} and $ \mathcal{H}_\Delta$ in Eq.~\eqref{eq-ham-delta}.
And the averaged boundary SP is defined as,
\begin{align}
\tilde{M}_z = \frac{1}{N_l}\sum_{\vec{l}} \sum_{E_n}  \langle E_n(\vec{l}) \vert \hat{P}_e^\dagger  \sigma_z(\vec{l}) \hat{P}_e \vert E_n(\vec{l})\rangle_{BdG},
\end{align}
where $\vec{l}$ is the ``edge coordinate'' on a $N_x\times N_y$ rectangular lattice and $\vec{l}\in\{(1,i_y),(i_x,N_y),(N_x,i_y),(i_x,1)\}$ with total sites $N_l=2N_x+2N_y-4$, $\hat{P}_e$ is the projection operator into the particle subspace, and $\vert E_n\rangle$ is solution of the BdG equation $\mathcal{H}_{BdG}\vert E_n\rangle = E_n \vert E_n\rangle$.

Next, we take the $d+ip$ pairing states in Eq.~\eqref{eq-sip-and-dip} for an example. 
And the numerical result is shown in Fig.~\ref{fig2}. 
It confirms that the averaged spontaneous SP $\tilde{M}_z=0$ corresponding to SOC strength $\alpha=0$ and it increases as increasing the $\alpha$, consistent with the analytical analysis for $\gamma_2^z$ in Eq.~\eqref{eq-gamma2z-approx}.
Moreover, similar results are found in three different cases: $\Delta_t < \Delta_s$, $\Delta_t = \Delta_s$ and $\Delta_t > \Delta_s$.
From this tendency, one may roughly estimate the induced SP $\tilde{M}_z\sim 0.02$ meV for Zr$_3$Ir, where the band splitting near the Fermi level due to SOC is about $\alpha k_F\sim100$ meV, $T_c\approx 2.3$ K and $\Delta_s \approx \Delta_t \sim 0.2$ meV~\cite{Shang2020b}.
Hence, we argue that the SP induced by TRS breaking unitary pairing is large enough to be detected by the $\mu$SR measurement in Zr$_3$Ir \cite{Shang2020b}.

\begin{figure}[t]
	\centering
	\includegraphics[width=0.8\linewidth]{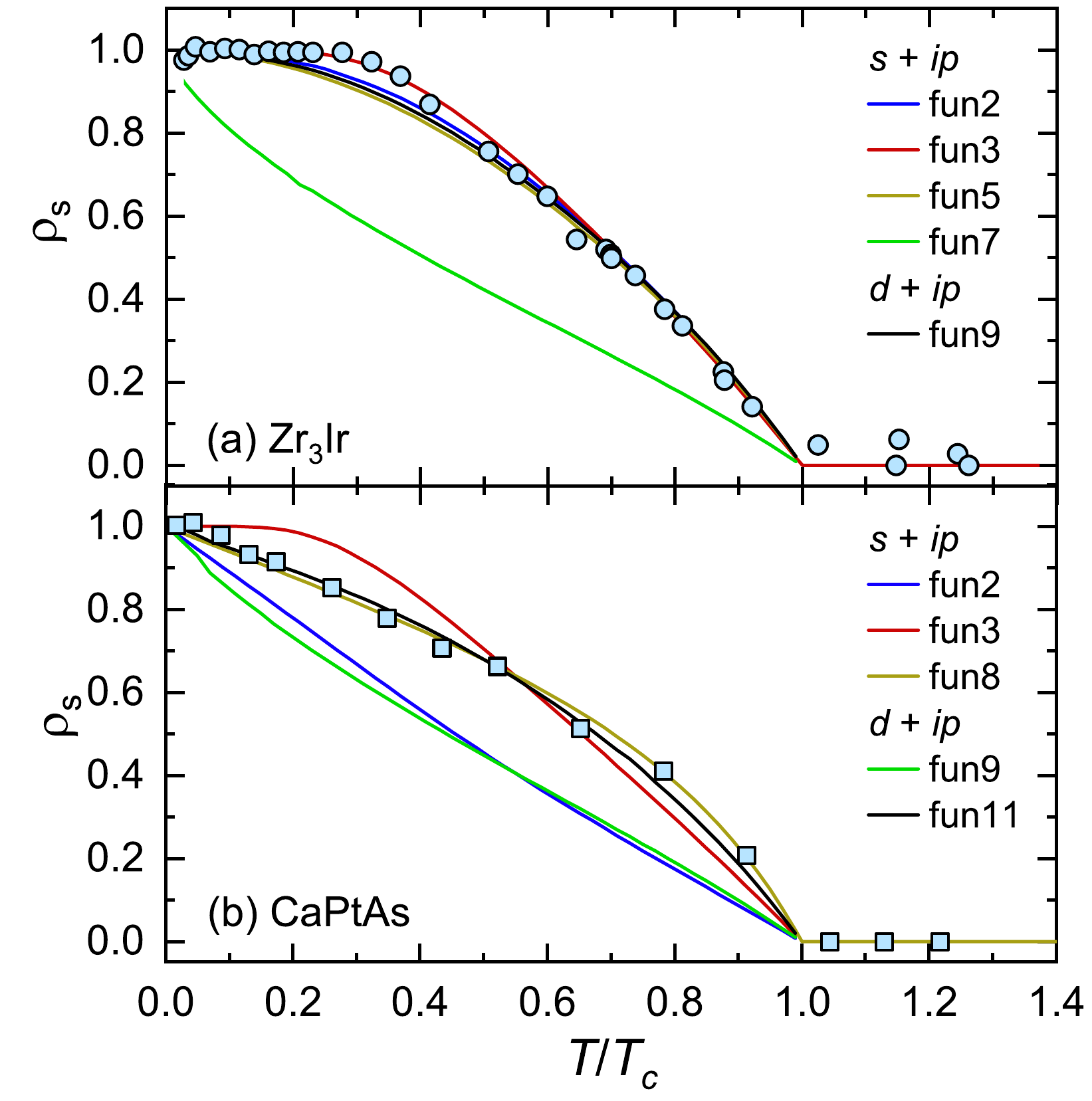}
	\caption{The superfluid density $\rho_s$ versus the reduced temperature $T$/$T_c$ for the noncentrosymmetric SCs. 
		(a) Zr$_3$Ir with a weak SOC and (b) CaPtAs with a strong SOC. The experimental data determined by the transverse-field muon-spin rotation measurements were taken from Ref.~\onlinecite{Shang2020a,Shang2020b}.
		The solid lines represent the $\rho_s$ fitted by different pairing functions. For CaPtAs, it might be also fitted by pairing fun\,5, 7, 10, and 12 (see Fig.~S1 in the Supplementary Materials).  }
	\label{fig3}
\end{figure}

\textit{Theoretical applications to noncentrosymmetric SCs.--}
Now, we discuss the compatibility of the above superconducting pairings with the real materials. 
Here we consider two different noncentrosymmetric SCs, Zr$_3$Ir \cite{Shang2020b} and CaPtAs \cite{Shang2020a}, both exhibiting a breaking of TRS in their superconducting states. 
In general, the admixture of singlet and triplet pairings is mostly determined by the strength of SOC. 
The energy scale of band splitting caused by SOC ($\alpha k_F/k_BT_c$), is expected to be much larger in CaPtAs than that in Zr$_3$Ir. 
Therefore, the Zr$_3$Ir exhibits a fully gapped SC, while gap nodes were observed in CaPtAs, suggesting a large triplet component in the latter case as the SOC increases. 

We first discuss the weak SOC case, i.e., Zr$_3$Ir. 
The mixed pairings in Eq.~\eqref{eq-sip-and-dip} give rise to the 9 gap functions by considering three limits: $\Delta_s\ll\Delta_t$, $\Delta_s\sim\Delta_t$ and $\Delta_s\gg\Delta_t$ (see Sec.~C in the Supplementary Materials).
If $\Delta_s$ $\gg$ $\Delta_t$, the gap functions become singlet, inconsistent with the broken TRS in Zr$_3$Ir. Therefore, we focus on the gap functions with $\Delta_s$ $\ll$ $\Delta_t$ or $\Delta_s$ $\sim$ $\Delta_t$. For $\Delta_s$ $\ll$ $\Delta_t$, triplet is dominated, the pairing function $\Delta_{s+ip}$ = $\Delta_{d+ip}$ $\sim$ 
$\Delta_0$$\lvert$Sin$\theta$$\lvert$
exhibits gap nodes, $\Delta_0$ being the superconducting gap at zero temperature whose value is also related to the magnitude of $\Delta_t$. 
Indeed, as shown by the fun\,2 in Fig.~\ref{fig3}(a), the calculated superfluid density $\rho_s$ shows a clear temperature dependence at low temperatures (see details of the calculations in Ref.~\onlinecite{Maisuradze2009}).
While for $\Delta_s$ $\sim$ $\Delta_t$, as shown in Fig.~\ref{fig3}(a), all fun\,5 and fun\,7 for $s+ip$ and fun\,9 for $d+ip$ exhibit temperature-dependent superfluid density, similar to the case of fun\,2. 
As for fun\,3 ($\Delta_{s+ip}$ $\sim$ $\Delta_0$$\sqrt{1 + \mathrm{Sin}^2\theta}$), the calculated $\rho_s$ is temperature-independent at low-$T$, an indication of fully-gapped SC, and shows remarkably good agreement with the experimental data. 
Therefore, the $s+ip$-pairing with $\Delta_s$ $\sim$ $\Delta_t$ might be applied to the noncentrosymmetric Zr$_3$Ir SC with weak SOC.

While for the strong SOC case (i.e., CaPtAs), the mixed pairings in Eq.~\eqref{eq-sip-and-dip} lead to 12 gap functions in total for the three different limits ($\Delta_s\ll\Delta_t$, $\Delta_s\sim\Delta_t$ and $\Delta_s\gg\Delta_t$) (see Sec.~C in the Supplementary Materials).
Considering the presence of both broken TRS and superconducting gap nodes in CaPtAs, we exclude the pairing functions with $\Delta_s$ $\gg$ $\Delta_t$. 
In the case of $\Delta_s$ $\sim$ $\Delta_t$, for $s+ip$ pairing, all fun\,3, fun\,5, and fun\,7 show a poor agreement with the experimental data (see Fig.~S1 in Supplementary Materials).
Then, we focus on the gap functions with $\Delta_s$ $\ll$ $\Delta_t$. 
Both fun\,2 (see Fig.~\ref{fig3}(b) for $s+ip$) and fun\,10 (see Fig.~S1 for $d+ip$) exhibit a strong temperature-dependent superfluid density, and deviate significantly from the experimental data. 
However, for fun\,8 ($\Delta_{s+ip}$ $\sim$ $\Delta_0$$\lvert\mathrm{Cos}\varphi\mathrm{Sin}\varphi\mathrm{Sin}\theta\lvert$),
the calculated $\rho_s$ shows a good agreement with the experimental data over the entire temperature range. 
Furthermore, the $\rho_s$ calculated from fun\,11 and fun\,12 for $d+ip$ are also highly consistent with the experimental data, and the fitting result of fun\,11 ($\Delta_{d+ip}$ $\sim$ $\Delta_0$$\lvert\mathrm{Cos}\varphi\mathrm{Sin}^2\theta - \mathrm{Sin}\theta + \tfrac{\mathrm{Sin}\theta}{\lvert\mathrm{Sin}\theta\lvert}\lvert$) is illustrated in Fig.~\ref{fig3}(b).
Therefore, both $s+ip$- and $d+ip$-pairings with $\Delta_s$ $\ll$ $\Delta_t$ might be applied to the noncentrosymmetric CaPtAs SC with strong SOC.
By comparing with the experimental data, we demonstrate that our theory might be practicable to both nodal and nodeless superconductivity with broken TRS in the noncentrosymmetric SCs.

\textit{Conclusion and discussion--.}
We briefly discuss the effects of the boundary SP on first-order (second-order) 2D topological SCs, which host topological Majorana edge (corner) states (see Sec.~D in the Supplementary Materials).
A purely helical $p$-wave SC supports a pair of helical Majorana edge modes protected by TRS \cite{qi2009,tanaka2009,sato2009}, which become fully gapped on each boundary for the $s+ip$ pairing states.
Furthermore, a second-order topological SC is achieved \cite{Benalcazar2017,langbehn2017,zhu2018,liu2018,hsu2018,wang2018a,Wang2018b,yan2018,pan2019,zhang2019a,zhang2019b,wu2019a,wu2019b,wu2020a,vu2020,zhang2020,wu2020b,wu2020c,laubscher_prb_2020,plekhanov_prb_2021} for the $d+ip$ case, which supports topological Majorana corner states (MCS).
They are kind of Jackiw-Rebbi zero modes \cite{Jackiw} sitting on each corner, protected by the combined $\sigma_d\mathcal{T}$ symmetry. Interestingly, we find that the boundary SP enlarges the edge gap to protect the MCSs.

In sum, we find that the TRS-breaking unitary pairing states could induce the spontaneous SP with the help of SOC, in the absence of external magnetic fields or Zeeman fields.
We propose that both $s+ip$ and $d+ip$ spontaneously break TRS and give rise to SP, which is induced to be perpendicular to both the real spin-triplet $\vec{d}$-vector and the SOC $\vec{g}$-vector. 
The averaged boundary SP is also estimated $\sim0.02$ meV for noncentrosymmetric SC, Zr$_3$Ir, which should be able to be detected in experiments.
Moreover, our theory can quantitatively describe the superfluid density of Zr$_3$Ir and CaPtAs noncentrosymmetric SCs.
Our result provides an alternative explanation to the TRS breaking, beyond the current understanding of such phenomena in the noncentrosymmetric superconductors.
We also notice a recent theoretical work demonstrating that the pairing symmetry might be $d+ip$ for Sr$_2$RuO$_4$~\cite{scaffidi2020}. 
Our theory may be also valid near an interface where spin-orbit coupling appears to explain the observations of broken TRS~\cite{luke1998,xia2006}.

\textit{Acknowledgments--.}
L.-H. Hu is indebted to W.~Yang, R.-X.~Zhang, C.~J.~Wu, and D.~F.~Agterberg for helpful discussions.	 
The work is initialized at the University of California, San Diego.
L.-H. Hu acknowledge the support of the Office of Naval Research (Grant No. N00014-18-1-2793) and Kaufman New Initiative research Grant No. KA2018-98553 of the Pittsburgh Foundation.
T. Shang acknowledge the support from the Natural Science Foundation of Shanghai (Grant No. 21ZR1420500).

\bibliographystyle{apsrev4-2}
\bibliography{ref}

\clearpage
\appendix

\setcounter{equation}{0}
\setcounter{figure}{0}
\setcounter{table}{0}
\setcounter{page}{1}

\renewcommand{\figurename}{FIG.\ S\!\!}
\renewcommand{\tablename}{Table \ S\!\!}

\begin{widetext}
\begin{center}
	\bf	Supplementary materials for ``Spontaneous magnetization in time-reversal symmetry-breaking unitary superconductors''
\end{center}
It contains the discussion for the TRS breaking unitary pairing states, the character table for $C_{4v}$ point group, the pairing functions for $s+ip$ and $d+ip$ with small/large SOC, and the effects of the spin-polarization on topological SCs.

\section{A. The TRS breaking unitary pairing states}
In this section, we discuss if the superconductivity-induced spin-polarization/magnetism ($\vec{M}\neq0$) will affect the relative phase difference between $\Delta_s$ and $\Delta_t$. The answer is ``Not'', and the reason is shown in the following.
Without loss of generality, we set the three order parameters as $\Delta_s$, $e^{i\theta} \Delta_t$ and $\vec{M}=(0,0,M_z)$, where $\Delta_s$, $\Delta_t$, $\theta$, $M_z$ are all real. 
In terms of order parameters $\Delta_s$, $\Delta_t$, $\theta$ and $M_z$, the free energy in Eq.~\eqref{eq-GL-free-energy} becomes,
\begin{align}
\mathcal{F}&=\alpha_s \Delta_s^2 + \alpha_t \Delta_t^2 + \alpha_M M_z^2 + \beta_s \Delta_s^4 + \beta_t \Delta_t^4
+ \beta_{st}  \Delta_s^2 \Delta_t^2 + 2\gamma_1 \cos 2\theta \Delta_s^2 \Delta_t^2 + 2\gamma_{2z}  \sin\theta M_z \Delta_s \Delta_t,
\end{align}
where $\alpha_s<0$, $\alpha_t<0$, $\alpha_M>0$, $\gamma_1>0$ and all others are positive. To minimize the above free energy, we have totally four equations,
\begin{align}
 \frac{\partial\mathcal{F}}{\partial\Delta_s} &= 2\alpha_s \Delta_s + 4\beta_s \Delta_s^3 + 2\beta_{st} \Delta_s \Delta_t^2 + 4\gamma_1 \cos 2\theta \Delta_s \Delta_t^2 + 2\gamma_{2z} \sin\theta M_z \Delta_t=0, \\
 \frac{\partial\mathcal{F}}{\partial\Delta_t} &= 2\alpha_t \Delta_t + 4\beta_t \Delta_t^3 + 2\beta_{st} \Delta_t \Delta_s^2 + 4\gamma_1 \cos 2\theta \Delta_t \Delta_s^2 + 2\gamma_{2z} \sin\theta M_z \Delta_s=0, \\
 \frac{\partial\mathcal{F}}{\partial M_z} &= 2\alpha_M M_z + 2\gamma_{2z} \sin\theta \Delta_s \Delta_t=0, \\
 \frac{\partial\mathcal{F}}{\partial \theta} &= -4\gamma_1 \sin 2\theta  \Delta_s^2 \Delta_t^2 + 2\gamma_{2z} \cos\theta  M_z \Delta_s \Delta_t=0.
\end{align}
Please notice that there is no intrinsic ferro-magnetism in this system ($\alpha_M>0$). 
By solving the third equation, we find that $M_z=-\gamma_{2z}/\alpha_M \sin\theta \Delta_s \Delta_t$, and substituting it into the fourth equation, we then have
\begin{align}
-4\gamma_1  \sin 2\theta \Delta_s^2 \Delta_t^2 - 2\gamma_{2z}  \cos\theta (\gamma_{2z}/\alpha_M \sin\theta  \Delta_s \Delta_t ) \Delta_s \Delta_t=0,
\end{align}
which leads to $\sin 2\theta=$0 so that $\theta=\pm\pi/2$ with the fact $\Delta_s\neq0$ and $\Delta_t\neq0$. 
After substituting $M_z$ and $\theta$ back into the first two equations, one could get the solutions for $\Delta_s$ and $\Delta_t$.

In summary, the time-reversal symmetry breaking superconductivity-induced spin-polarization/magnetism will not make the relative phase difference between $\Delta_s$ and $\Delta_t$ away from $\pm\pi/2$. In other words, the unitary pairing state $\Delta_s\pm i\Delta_t$ would not be affected by the induced magnetism.

\section{B. Character table for $C_{4v}$ point group}
In the main text, we consider the $C_{4v}$ point group to discuss the representation of SP and superconducting order parameters.
This group is generated by three independent symmetry operators ($\sigma_v$, $\sigma_d$ and $R_{4z}$).   
$\sigma_v$ are the vertical reflection planes along $x$ and $y$, $\sigma_d$ are the diagonal reflection planes along the $x\pm y$ lines, and $R_{4z}$ is the four-fold rotation along $z$ axis. 
The character table is shown as follows (see Fig.~\ref{tab-c4v}).
\begin{table}[!htbp]
	\begin{ruledtabular}
		\begin{tabular}{c|c|c|c}
			Orders    & Mirror $\sigma_v$   & Mirror $\sigma_d$ & $R_{4z}$     \\ \cline{1-4}
			s-wave    & $+$  & $+$        & $+$      \\ 
			d-wave    & $+$  & $-$        & $-$      \\ \hline \hline
			$\vec{d}_{A_{1}}$      & $+$  & $+$      & $+$      \\ 
			$\vec{d}_{A_{2}}$      & $-$  & $-$      & $+$      \\ 
			$\vec{d}_{B_{1}}$      & $+$  & $-$      & $-$     \\ 
			$\vec{d}_{B_{2}}$      & $-$  & $+$      & $-$     \\ \hline\hline
			$M_{A_{2}}$         & $-$  & $-$      & $+$     \\ 
		\end{tabular}
	\end{ruledtabular}
	\caption{The character table for all orders (spin-single/triplet pairings and spin-polarization) based on the $C_{4v}$ point group. The independent symmetry operators involve the mirror $\sigma_v$, $\sigma_d$ and the four-fold rotation $R_{4z}$.}
	\label{tab-c4v}
\end{table}

\section{C. Pairing functions for $s+ip$ and $d+ip$}
In the main text, we propose two mixed pairing states that can spontaneously induce SP with the help of SOC. They are given by Eq.~(7) in the main text,
\begin{align}
\begin{split}
\hat{\Delta}_{s+ip} &= \left(\Delta_s \Psi_s(\mathbf{k}) + i\Delta_t \vec{d}_{A_{2}}(\mathbf{k})\cdot\vec{\sigma} \right)i\sigma_y,\\
\hat{\Delta}_{d+ip} &= \left(\Delta_s \Psi_d(\mathbf{k}) + i\Delta_t \vec{d}_{B_{2}}(\mathbf{k})\cdot\vec{\sigma} \right)i\sigma_y.
\end{split}
\end{align}
To apply our theory to realistic noncentrosymmetric SCs, we list the normalized gap functions on the Fermi surfaces. Here we consider both weak SOC and large SOC cases,
\begin{itemize}
	\item[(I)] Weak SOC case. \\
	\begin{align}
	\begin{split}
	\Delta_{s+ip} &= \sqrt{(\Delta_s\Psi_s(k)^2 + \Delta_t^2(k_x^2 + k_y^2)},\\
	\Delta_{d+ip} &= \sqrt{\Delta_s^2(k_x^2 - k_y^2)^2 + \Delta_t^2(k_x^2 + k_y^2)}.
	\end{split}
	\end{align}
	\item[(II)] Large SOC case. Here we ignore the inter-band pairing for simplicity.\\
	For TRS breaking unitary $s+ip$ pairing states,
	\begin{align}
	\begin{split}
	\Delta_{s+ip}(1) &= \lvert \Delta_s\Psi_s(k)\frac{k_x + ik_y}{\sqrt{k_x^2 + k_y^2} } +  \Delta_t\Psi_1(k) \lvert,\\
	\Delta_{s+ip}(2) &= \lvert -\Delta_s\Psi_s(k)\frac{k_x - ik_y}{\sqrt{k_x^2 + k_y^2} } +  \Delta_t\Psi_2(k) \lvert,	
	\end{split}
	\end{align}
	where $\Psi_1(k)=(-k_x+ik_y)+\tfrac{(k_x+ik_y)^3}{k_x^2+k_y^2} $ and  $\Psi_2(k)=(k_x+ik_y)-\tfrac{(k_x-ik_y)^3}{k_x^2+k_y^2} $.
	
	For TRS breaking unitary $d+ip$ pairing states,
	\begin{align}
	\begin{split}
	\Delta_{d+ip}(1) &= \lvert \Delta_s(k_x^2 - k_y^2)\frac{k_x + ik_y}{\sqrt{k_x^2 + k_y^2} } +  \Delta_t\Psi_1(k) \lvert,\\
	\Delta_{d+ip}(2) &= \lvert -\Delta_s(k_x^2 - k_y^2)\frac{k_x - ik_y}{\sqrt{k_x^2 + k_y^2} } +  \Delta_t\Psi_2(k) \lvert.	
	\end{split}
	\end{align}
	where $\Psi_1(k)=(-k_x-ik_y)+k_F\tfrac{k_x+ik_y}{\sqrt{k_x^2+k_y^2}} $ and  $\Psi_2(k)=(k_x-ik_y)-k_F\tfrac{k_x-ik_y}{\sqrt{k_x^2+k_y^2}} $.
\end{itemize}
Please note that the angular form factors for spin-singlet pairing could be either s-wave ($\psi_s(\mathbf{k})\sim \{1,\cos k_x +\cos k_y,\cos k_x\cos k_y\}$) or d-wave ($\psi_d(\mathbf{k})\sim \cos k_x - \cos k_y$).
As for p-wave pairing symmetries, the in-plane $\vec{d}$ vectors used for $s+ip$ and $d+ip$ are, 
\begin{align}
\begin{split}
\text{for } s+ip\text{ case: }  &\vec{d}_{A_{2}}(\mathbf{k}) = (k_x,k_y), \\
\text{for } d+ip\text{ case: }  &\vec{d}_{B_{2}}(\mathbf{k}) = (k_x,-k_y). 
\end{split}
\end{align}

\subsection{(1) Gap functions $\Delta(k_x,k_y)$ for $s+ip$ and $d+ip$ with weak SOC}
\begin{itemize}
\item[] {\bf Fun 1:} $\Delta_t\ll \Delta_s$,  $\Delta(k_x,k_y)\sim \Delta_0$. 
\item[] {\bf Fun 2:} $\Delta_t\gg \Delta_s$,  $\Delta(k_x,k_y)\sim \Delta_0\vert \sin\theta\vert$. 
\item[] {\bf Fun 3:} $\Delta_t\sim \Delta_s$, $\Delta(k_x,k_y)\sim \Delta_0\sqrt{1+\sin^2\theta}$.  
\item[] {\bf Fun 4:} $\Delta_t\ll \Delta_s$,  $\Delta(k_x,k_y)\sim \Delta_0 \sin^2\theta$.  
\item[] {\bf Fun 5:} $\Delta_t\sim \Delta_s$, $\Delta(k_x,k_y)\sim \Delta_0\sqrt{1+\sin^2\theta}$.  
\item[] {\bf Fun 6:} $\Delta_t\ll \Delta_s$,  $\Delta(k_x,k_y)\sim \Delta_0 \vert [-1+2\cos^2\varphi\sin^2\theta][-1+2\sin^2\theta\sin^2\varphi] \vert $.
\item[] {\bf Fun 7:} $\Delta_t\sim \Delta_s$, $\Delta(k_x,k_y)\sim \Delta_0 \sqrt{\sin^2\theta + \{[-1+2\cos^2\varphi\sin^2\theta][-1+2\sin^2\varphi\sin^2\theta]\}^2 }$.
\item[] {\bf Fun 8:} $\Delta_t\ll \Delta_s$,  $\Delta(k_x,k_y)\sim \Delta_0\sin^2\theta\vert\cos2\varphi\vert$. 
\item[] {\bf Fun 9:} $\Delta_t\sim \Delta_s$, $\Delta(k_x,k_y)\sim \Delta_0\vert\sin\theta\vert \sqrt{1+\sin^2\theta\cos^22\varphi}$. 
\end{itemize}
These 9 gap functions are used to fit the superfluid density $\rho_s$ in the main text for Zr$_3$Ir, see Fig.~3(a).

\subsection{(2) Gap functions for $s+ip$ and $d+ip$ with large SOC}
\begin{itemize}
\item[] {\bf Fun 1:}  $\Delta_t\ll \Delta_s$,  $\Delta(k_x,k_y)\sim \Delta_0$. 
\item[] {\bf Fun 2:}  $\Delta_t\gg \Delta_s$,  $\Delta(k_x,k_y)\sim \Delta_0\vert \sin\theta\cos2\varphi\vert$. 
\item[] {\bf Fun 3:}  $\Delta_t\sim \Delta_s$, $\Delta(k_x,k_y)\sim \Delta_0\vert 1 + 2i\sin\theta\sin2\varphi \vert$.
\item[] {\bf Fun 4:}  $\Delta_t\ll \Delta_s$,  $\Delta(k_x,k_y)\sim \Delta_0\sin^2\theta$.  
\item[] {\bf Fun 5:}  $\Delta_t\sim \Delta_s$, $\Delta(k_x,k_y)\sim \Delta_0\vert \sin^2\theta + 2i\sin\theta\sin 2\varphi \vert$. 
\item[] {\bf Fun 6:}  $\Delta_t\ll \Delta_s$,  $\Delta(k_x,k_y)\sim \Delta_0\vert (-1+2\cos^2\varphi\sin^2\theta)(-1+2\sin^2\theta\sin^2\varphi) \vert$. 
\item[] {\bf Fun 7:}  $\Delta_t\sim \Delta_s$, $\Delta(k_x,k_y)\sim \Delta_0\vert (-1+2\cos^2\varphi\sin^2\theta)(-1+2\sin^2\theta\sin^2\varphi) +2i\sin\theta\sin2\varphi  \vert$. 
\item[] {\bf Fun 8:}  $\Delta_t\gg \Delta_s$,  $\Delta(k_x,k_y)\sim \Delta_0\vert \sin2\varphi\sin\theta \vert$. 
\item[] {\bf Fun 9:}  $\Delta_t\ll \Delta_s$,  $\Delta(k_x,k_y)\sim \Delta_0\vert \cos2\varphi \sin^2\theta \vert$. 
\item[] {\bf Fun 10:} $\Delta_t\gg \Delta_s$,  $\Delta(k_x,k_y)\sim \Delta_0\vert -\sin\theta + \frac{\sin\theta}{\vert\sin\theta\vert} \vert $. 
\item[] {\bf Fun 11:} $\Delta_t\sim \Delta_s$, $\Delta(k_x,k_y)\sim \Delta_0 \vert \cos2\varphi\sin^2\theta + (-\sin\theta + \frac{\sin\theta}{\vert\sin\theta\vert}) \vert$. 
\item[] {\bf Fun 12:} $\Delta_t\sim \Delta_s$, $\Delta(k_x,k_y)\sim \Delta_0 \vert \cos2\varphi\sin^2\theta - (-\sin\theta + \frac{\sin\theta}{\vert\sin\theta\vert}) \vert$. 
\end{itemize}
These 12 gap functions are used to fit the superfluid density $\rho_s$ in the main text for CaPtAs, see Fig.~3(b).

\begin{figure}[t]
	\centering
	\includegraphics[width=0.5\linewidth]{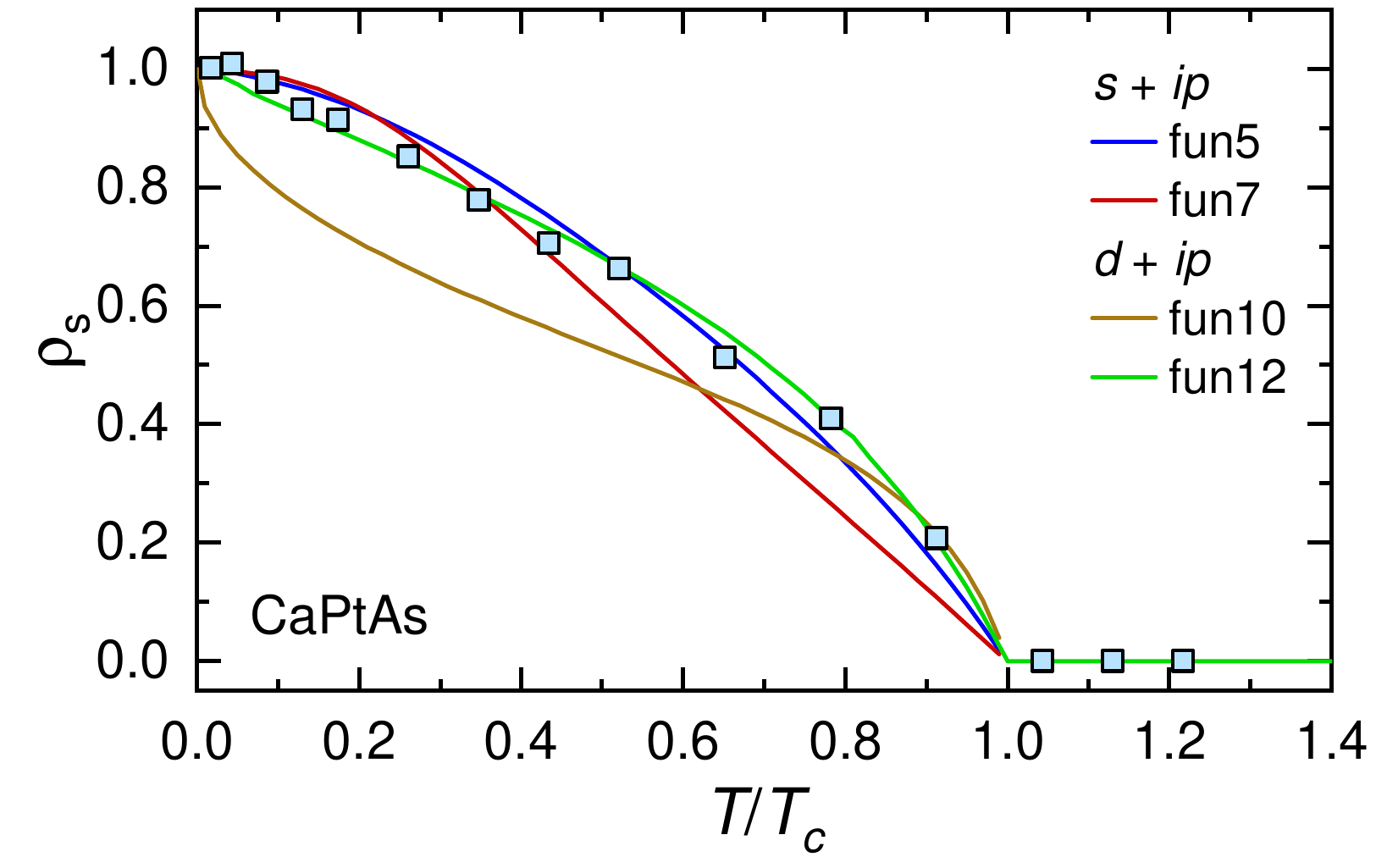}
	\caption{The superfluid density $\rho_s$ versus the reduced temperature $T$/$T_c$ for the noncentrosymmetric CaPtAs superconductor. The experimental data determined by the transverse-field muon-spin rotation measurements were taken from Ref.~\onlinecite{Shang2020a}, while the solid lines represent the $\rho_s$ calculated using different pairing functions.}
		\label{figs1}
\end{figure}

\section{D. Effects of the spin-polarization on topological superconductors}
It is also worthwhile to study effects of the boundary SP on first-order (second-order) topological superconductors, which host topological Majorana edge (corner) states.
From the viewpoint of symmetry arguments, we learn that two singlet-triplet mixed gap functions $\Delta_{s+ip}$ and $\Delta_{d+ip}$ in Eq.~\eqref{eq-sip-and-dip} not only break TRS but also lead to the boundary SP with the help of the Rashba-type SOC.
We firstly discuss the $s+ip$ pairing states.
It is well known that a purely p-wave superconductor supports a pair of helical Majorana edge modes (HMEM) protected by TRS \cite{qi2009,tanaka2009,sato2009}, which is descried by the edge Hamiltonian $\mathcal{H}_{edge}(\vec{l}) = A(\vec{l}) k_{l} s_z$. 
Here $k_{l}$ is the momentum on edge $l$ and $\vec{s}$ are the Pauli matrices defined for the HMEM basis.
Both the s-wave pairing potential and the boundary SP are the sources for the mass generation to the HMEMs, resulting in the edge Hamiltonian as, 
\begin{align}
\mathcal{H}_{edge}(\vec{l}) = A(\vec{l}) k_{l} s_z + (M_s(\vec{l})+M_{sp}(\vec{l}))s_x,
\label{hgen}
\end{align}
with $M_s(\vec{l})M_{sp}(\vec{l})>0$, thus the HMEMs are fully gapped on each boundary with the same mass sign as expected.

As for the $d+ip$ case, a second-order topological superconductor (HOTSC) is achieved \cite{Benalcazar2017,langbehn2017,zhu2018,liu2018,hsu2018,wang2018a,Wang2018b,yan2018,pan2019,zhang2019a,zhang2019b,wu2019a,wu2019b,wu2020a,vu2020,zhang2020,wu2020b,wu2020c,laubscher_prb_2020,plekhanov_prb_2021}, which supports topological Majorana corner states (MCS).
Since both the d-wave gap function and the boundary SP $M_z$ are odd under $\sigma_d$, they serve as staggered mass potentials for the HMEMs in the edge theory.
Once the staggering masses are obstructed, the Jackiw-Rebbi zero modes \cite{Jackiw} appear on each corner, protected by the combined $\sigma_d\mathcal{T}$ symmetry.
To show this precisely, we construct the edge Hamiltonian in the spirit of $k\cdot p$ theory and take the $l_1$-edge for an example. 
Firstly, we solve $\mathcal{H}_{BdG}(i\partial_x,k_y=0)\vert \chi_\pm \rangle = 0$ for the HMEMs basis $(\vert\chi_{1}\rangle,\vert \chi_{-1}\rangle)$ for $x\ge0$.
After some algebra, we find $\ket{\chi_{1}}=(e^{i \theta}\ket{p_{1}}+e^{-i \theta}\ket{h_{1}}) e^{-\xi x} /N_0$ with $\tan\theta=(\alpha + \Delta_t)/\sqrt{\Delta_t^2-\alpha^2}$ and $N_0$ is the normalization constant. 
Here we set $\vert p_1\rangle=(0,0,-i,1)^T/\sqrt{2}$ and $\vert h_1\rangle=(i,1,0,0)^T/\sqrt{2}$.
The TRS-partner satisfies $\vert \chi_{-1}\rangle=\mathcal{T}\vert \chi_{1}\rangle$.
And the localization length $\xi$ satisfies $(\xi^2+\mu)-\sqrt{\Delta_t^2-\alpha^2} \xi =0$, which gives rise to the criterion for the helical Majorana states to exist, $\Delta_t^2-\alpha^2>0$.
Therefore, we obtain the edge Hamiltonian $\mathcal{H}_{l_1}(k_y)$,
\begin{align}
\mathcal{H}_{l_1}(k_y)=A(l_1)k_y s_x + (M_s(l_1)+M_{sp}(l_1))s_y,
\label{hfin}
\end{align}
where $A(l_1)=-(\Delta_t-\alpha \cos 2\theta)$, $M_s(l_1)=-\Delta_d/2$, and $M_{sp}(l_1)=M_z \cos 2\theta$.
We also find that $M_s(l_1)M_{sp}(l_2)>0$ as expected.
Note that $M_{sp}=0$ for the $\alpha=0$ case, which is consistent with the GL theory. 
Likewise, the mass terms for other edges are,
\begin{align}
M_{s}(l_i)=\left\{
\begin{array}{rcl}
\Delta_d/2     &      \text{ for even i},\\
-\Delta_d/2   &      \text{ for odd i}.
\end{array} \right.
\end{align}
and $M_s(l_i)M_{sp}(l_i)>0$.
It indicates that the boundary SP enlarges the edge gap to protect the MCSs.
We notice that the quadrupolar structure of the SP serves as a signature of a 2D HOTSC with an external applied Zeeman field \cite{plekhanov_prb_2021}.
However, in our work, the superconductivity-induced SP occurs in the absence of external magnetic fields or Zeeman fields.

\end{widetext}
\end{document}